\documentclass[11pt,a4paper]{article}

\usepackage{amsmath, amsthm,mathtools,setspace, hyperref, url,amssymb,upgreek,textgreek,slashed}

\usepackage[citation-order]{amsrefs}
\usepackage{graphicx}
\usepackage{color}
\usepackage{amsfonts}
\usepackage{amssymb}
\usepackage{amscd}
\usepackage{array}
\usepackage{subfigure}
\usepackage{xypic}
\usepackage[margin=2.5cm]{geometry}
\usepackage{booktabs}
\usepackage{hyperref}
\usepackage{longtable}
\usepackage{pdflscape}
\usepackage{colortbl}
\usepackage{enumitem}

\newcommand{\IP}{\mathbb{P}}

\newcommand{\cN}{{\mathcal N}}

\newcommand{\cA}{{\mathcal A}}

\newcommand{\cK}{{\mathcal K}}

\begin{document}

\begin{center}
{
{\bf\Large Intelligent Explorations of the String Theory Landscape\footnote{Chapter prepared for the Wold Scientific volume {\itshape Machine-learning in Theoretical Physics and Pure Mathematics.}}}\\[12pt]
\vspace{2mm}
\normalsize
{{\large Andrei Constantin}$^{}$\footnote{andrei.constantin@physics.ox.ac.uk}
} 
\bigskip}\\[0pt]
\vspace{0.12cm}
{\it 
Rudolf Peierls Centre for Theoretical Physics, University of Oxford\\[-2pt]
Parks Road, Oxford OX1 3PU, UK\\[2pt]
Wolfson College, Linton Road, Oxford, UK
}\\[2ex]
\end{center}

\begin{center}
{\bfseries\large Abstract}
\vspace{8pt}

{\noindent
\begin{minipage}{16cm}
\small
The goal of identifying the Standard Model of particle physics and its extensions within string theory has been one of the principal driving forces in string phenomenology. Recently, the incorporation of artificial intelligence in string theory and certain theoretical advancements have brought to light unexpected solutions to mathematical hurdles that have so far hindered progress in this direction. In this review we focus on model building efforts in the context of the $E_8\times E_8$ heterotic string compactified on smooth Calabi-Yau threefolds and discuss several areas in which machine learning is expected to make a difference. 

\end{minipage}
}
\end{center}

\tableofcontents
\newpage
{\begin{flushright}
\itshape 
This paper is dedicated to the memory of Graham G.~Ross.
\end{flushright}
}

\section{Introduction}

Despite the wealth of settings available in string theory, it is currently not known how to embed the Standard Model of particle physics in any concrete string model. The primary reason for this is the sheer mathematical difficulty associated with the analysis of string compactifications. Numerous mathematical choices have to be made in order to specify a string compactification and the physical properties of the resulting four-dimensional quantum field theory depend on these choices in very intricate ways. 

Ideally, one would start with the empirical properties of the Standard Model and derive in a bottom-up fashion the topology and geometry of the underlying string compactification. Unfortunately, such a direct bottom-up approach has never been a real option. The reasons are multiple. On the one hand there are too many physical properties to account for in the Standard Model: the gauge group, the particle content, as well as a large number of free parameters such as the masses of the elementary particles and the strengths of the interaction couplings. On the other hand, these physical properties are related in a complicated way to the underlying topology and geometry: the particle spectrum is often computed in terms of cohomology groups, while the free parameters, which in principle can be dynamically traced back to the string length scale, depend on geometrical quantities that are difficult to find explicitly, such as the Calabi-Yau metric. On top of these complications there is the problem of moduli dependence: the compactification spaces come in infinite families, labelled by continuous parameters, which manifest as massless scalar fields in the low-energy theory. Finding mechanisms for dynamically fixing these parameters is non-trivial and in the absence of such mechanisms very little can be said about the quantitative properties of the low-energy theory. 

The alternative top-down approach to string phenomenology has only met with limited success. In this approach the internal topology and geometry are fixed at the start and the ensuing physical properties of the four-dimensional quantum field theory are subsequently derived. The difficulty here lies in the huge number of choices that can be made about the internal space -- the model building experience of the past few decades has taught us much about the magnitude of this problem and about how (and also about how not) to approach it. The first lesson is that the size of the string landscape is much larger than previously thought. The famous first estimate of $O(10^{500})$ consistent type IIB flux compactifications \cite{Douglas:2003um} seems rather conservative in comparison with the latest estimates. For instance, in Ref.~\cite{Taylor:2015xtz} it was shown that a single elliptically fibered four-fold gives rise to $O(10^{272,000})$ F-theory flux compactifications. The second lesson is that the number of compactifications that match the symmetry group and the particle spectrum of the Standard Model is very large, despite representing only a tiny fraction of all consistent compactifications to four dimensions. In Ref.~\cite{Constantin:2018xkj} it was argued that there are at least $10^{23}$ and very likely up to $10^{723}$ heterotic MSSMs, while the authors of Ref.~\cite{Cvetic:2019gnh} argued for the existence of a quadrillion standard models from F-theory. 
The third lesson is that these numbers are so large that traditional scanning methods cannot be used for systematic exploration. One can, of course, focus on small, accessible corners of the string landscape and this approach has been successful to some extent. For instance, in Refs.~\cite{Anderson:2013xka, Constantin:2018xkj}, some $10^7$ pairs of Calabi-Yau three-folds and holomorphic bundles leading to $SU(5)$ heterotic string models that can accommodate the correct MSSM spectrum have been explicitly constructed. 

Constructing effective field theories from string theory that agree with the Standard Model beyond the gauge group and the particle spectrum is non-trivial. 
On the one hand, the constraints that need to be imposed are mathematically and computationally challenging. On the other hand, even if these technical hurdles could somehow be resolved so as to include more constraints in the search algorithm, there is a high probability that no viable models would be found, unless the search space is considerably enlarged beyond the current possibilities. What is then needed is a tool-set of tailored search methods, that can quickly detect phenomenologically rich patches of the string landscape without systematically scanning over all compactifications and which can quickly implement a large number of checks that go beyond the usual spectrum considerations. 
The implementation of such methods is now being made possible through the emergence of new techniques of optimisation and search, in particular machine learning. In this sense the exploration of the string landscape in the search of familiar Physics is akin to the search for new Physics in the vast experimental data generated by present-day particle colliders, which also relies heavily on machine learning techniques. Experiment and theory need to converge and machine learning is likely to play a key role in bridging the gap between them. 

Machine learning essentially offers a mid-way alternative that avoids the difficulties inherent to both top-down and bottom-up approaches. While not solving directly for the ideal internal geometry and topology, methods such as reinforcement learning and genetic algorithms are capable of identifying many, and possibly all the viable models available within certain classes of compactifications after exploring only a tiny fraction of the entire range of possibilities \cite{Abel:2014xta, Halverson:2019tkf, Cole:2019enn, Larfors:2020ugo, Constantin:2021for, Krippendorf:2021uxu, Abel:2021rrj, Abel:2021ddu, Cole:2021nnt,Loges:2021hvn}. 

In the following discussion we will focus on model building efforts in the context of the $E_8\times E_8$ heterotic string compactified on smooth Calabi-Yau threefolds with holomorphic vector bundles. This has been the earliest and arguably the most promising proposal for connecting string theory to particle physics, but by no means the only one. Indeed, machine learning techniques have been successfully used in recent years in several other string theory contexts, starting with the early works of Refs.~\cite{He:2017set, He:2017aed, Ruehle:2017mzq, Carifio:2017bov, Krefl:2017yox} (see also the reviews~\cite{Ruehle:2020jrk, He:2020mgx}). 
Our discussion will focus on three propositions: 
\begin{itemize}\itemsep0em
\item[1.] A much larger portion of the heterotic string landscape can now be accessed through the use of heuristic methods of search.
\item[2.] The recent discovery of analytic formulae for bundle-valued cohomology has lead to a significant speed up in a number of checks that go beyond the net number of families.
\item[3.] The computation of physical couplings from string theory has been advanced by the development of machine learning algorithms for the numerical computation of Calabi-Yau metrics and hermitian Yang-Mills connections on holomorphic vector bundles.
\end{itemize}

In the following sections I will expand on these ideas, identifying a number of sub-problems where machine learning can make a difference.

\section{Heterotic String Model Building: an Overview}
In the heterotic string context, the problem of constructing a low-energy limit that recovers the Standard Model can be phrased as a two-step mathematical problem encoded by a pair $(X,V)$ consisting of a smooth, compact Calabi-Yau threefold $X$ and a slope-stable holomorphic vector bundle $V$ over $X$. The first step involves topology and algebraic geometry and concentrates on the identification of Calabi-Yau threefolds and holomorphic bundles with certain topological and quasi-topological properties. The second step involves differential geometry and concentrates on the problem of computing the Ricci-flat metric on the Calabi-Yau threefold, the hermitian Yang-Mills connection on the holomorphic vector bundle, as well as the harmonic representatives of certain bundle-valued cohomology classes that are in one-to-one correspondence with the low-energy particles. If achievable, these two steps would produce for every pair $(X,V)$ a class of four-dimensional effective field theories whose properties would be expressed in terms of the moduli determining the internal geometry. Fixing the moduli adds another layer of complication to the problem.

\subsection{Generalities}
The $E_8 \times E_8$ heterotic string theory has an in-built gauge symmetry, with each of the $E_8$ factors large enough to accommodate the Standard Model gauge group, as well as some of the standard GUT groups: $SU(5)$, $SO(10)$ and $E_6$. The two $E_8$ factors decouple at low energies: if the Standard Model gauge group is embedded in a single $E_8$, the other $E_8$ factor remains hidden and does not play a role in the initial construction of the low-energy theory and its particle spectrum. The hidden $E_8$ can, however, play an important role in moduli stabilisation.

At low energies, the $E_8\times E_8$ heterotic string theory in flat space can be consistently truncated to ten-dimensional $\mathcal N = 1$ supergravity coupled to $E_8\times E_8$ super-Yang-Mills theory. The gauge group and the multiplets of the Standard Model are naturally contained in the super-Yang-Mills theory. In order to make contact with empirical particle physics, one needs to dimensionally reduce the theory to four dimensions and to specify a non-vanishing background for the gauge fields, which has the double effect of partially breaking one of the $E_8$ factors and generating a chiral spectrum in four dimensions. Mathematically, one needs to specify a six-dimensional manifold~$X$ for the compactification space and a vector bundle $V$ over it whose connection specifies the background gauge fields. The gauge transformations available in four dimensions are the $E_8\times E_8$ transformations which commute with the internal gauge transformations. This implies that the unbroken subgroup of $E_8\times E_8$ is the commutant $H$ of the structure group $G$ of~$V$. The quantum numbers of the four-dimensional multiplets are determined by decomposing the adjoint representation of $E_8\times E_8$ under $G\times H$. The bundle $V$ decomposes into two parts, called the visible bundle and the hidden bundle, corresponding to the two $E_8$ factors. In order to obtain the usual GUT groups $SU(5)$, $SO(10)$ and $E_6$, the structure group of the visible bundle has to be $SU(5)$, $SU(4)$ or $SU(3)$, respectively. 

Often, the compactification data $(X,V)$ is chosen such that it leaves ${\mathcal N=1}$ supersymmetry unbroken in four dimensions at the compactification scale. On the one hand this choice simplifies the analysis, on the other hand it makes use of the advantages offered by ${\mathcal N=1}$ supersymmetry for Physics beyond the Standard Model, especially in combination with grand unification ideas. 
The implications of retaining ${\mathcal N=1}$ supersymmetry in four dimensions for the compactification data $(X,V)$ were first analysed in Ref.~\cite{Candelas:1985en}. If the structure of space-time is assumed to be $\mathbb R^4\times X$, where $\mathbb R^4$ is four-dimensional Minkowski space and $X$ is a compact six-dimensional manifold, the vanishing of the supersymmetry variation of the four-dimensional fields requires, in the simplest setting, that $X$ supports the existence of a covariantly constant spinor, which forces $X$ to be a Ricci-flat K\"ahler manifold. Finding Ricci-flat metrics on K\"ahler manifolds is a notoriously difficult problem, however, as Calabi conjectured \cite{Calabi1954, Calabi1957} and Yau later proved \cite{Yau:1977}, a simple topological condition on a K\"ahler manifold $X$, namely the vanishing $c_1(X)=0$ of its first Chern class, guarantees the existence of a unique Ricci-flat metric in each K\"ahler class. Such spaces are known as Calabi-Yau manifolds. The simple criterion offered by the Calabi-Yau theorem made possible the construction of large classes of examples, such as complete intersections in products of projective spaces \cite{Candelas:1987kf,Anderson:2015iia} (CICY threefolds for short), as well as hypersurfaces and complete intersections in toric varieties \cite{Kreuzer:2000xy}.

The requirement of $\cN=1$ supersymmetry in four dimensions also implies that the field strength on the vector bundle $V\rightarrow X$ satisfies the Hermitian Yang-Mills equations, $F_{ab}=F_{\bar a\bar b}=0$ and $g^{a\bar b}F_{a\bar b}=0$. These equations are difficult to solve explicitly, not least because they involve the Ricci-flat metric $g$ on $X$. Fortunately, there is a theorem due to Donaldson \cite{Donaldson:1985zz} (in complex dimension two), and Uhlenbeck and Yau \cite{uhlenbeck1986existence} (in arbitrary dimension), which proves that on a K\"ahler manifold, the Hermitian Yang-Mills equations admit a unique solution if and only if $V$ is holomorphic and slope-polystable. While in general it is non-trivial to check that a holomorphic bundle is polystable, for certain classes of bundles there exist algebro-geometric methods that make such checks more tractable (see Ref.~\cite{Anderson:2008ex}). 
Finally, the theory is anomaly-free if and only if $V$ and the tangent bundle $TX$ are related by the constraint $dH \sim tr(F\wedge F) - tr(R\wedge R)$, where $H$ is the field strength associated with the Kalb-Ramond 2-form $B$-field and $R$ is the curvature of $X$. The simplest solution to this constraint, known as the standard embedding, is to take the vector bundle $V$ to be the holomorphic tangent bundle $TX$, to set the gauge connection equal to the spin connection and $H = 0$.

\subsection{Three generation models} The initial heterotic model building efforts focused on the standard embedding and produced a handful of three-generation supersymmetric $E_6$ GUTs \cite{Greene:1986bm, Greene:1986jb,Schimmrigk:1987ke, Schimmrigk:1989ad, Braun:2009qy, Braun:2011ni}. A single multiplet in the fundamental ${\bf 27}$ representation of $E_6$ or the anti-fundamental $\overline{\bf 27}$ representation contains all the fermions in one family of the Standard Model. Since the number of ${\bf 27}$ multiplets is counted by the Hodge number~$h^{2,1}(X)$ and the number of $\overline{\bf 27}$-multiplets is given by the other non-trivial Hodge number~$h^{1,1}(X)$, in order to obtain three generations of quarks and leptons at low energies the threefold $X$ must satisfy
\begin{equation}
3 = |h^{2,1}(X) - h^{1,1}(X)| = \frac{1}{2}|\chi(X)|~,
\end{equation}
where $\chi(X)$ is the Euler characteristic of $X$ and the matching numbers of generations and anti-generations are assumed to pair up and acquire mass at a high energy scale.

The paucity of three generation standard embedding models obtained over the years is not unexpected. In fact what is remarkable is that any three-generation models at all could be found in this way. This is so because in order to break the $E_6$ symmetry down to $SU(3)\times SU(2)\times U(1)$, the manifold $X$ needs to be non-simply connected and, unfortunately, the number of known examples of non-simply connected Calabi-Yau threefolds with Euler characteristic equal to $\pm 6$ is very small (equal to $5$ according to the slightly old tabulation of Ref.~\cite{Candelas:2016fdy}). The requirement of non-simple connectedness comes from the fact that the standard GUT symmetry breaking mechanism makes use of the existence of topologically non-trivial gauge fields with vanishing field strengths (Wilson lines) on manifolds with non-trivial fundamental group. Note that since their field strengths vanishes, the Wilson lines do not contribute to the Hermitian Yang-Mills equations nor to the anomaly cancellation condition, so no additional complications arise.

The realisation that more general vector bundles on Calabi-Yau threefolds provide true solutions of the heterotic string opened up a much wider class of compactifications in which one could also construct $SO(10)$ and $SU(5)$ GUTs \cite{Distler:1986wm, Distler:1987ee}. While the number of available choices for $X$ remains relatively small due to the requirement of non-simply connectedness \cite{Candelas:2008wb, Braun:2010vc, Candelas:2010ve, Braun:2017juz, Candelas:2015amz, Candelas:2016fdy, Larfors:2020weh}, the number of possibilities for $V$ is virtually unbounded. Various constructions of holomorphic stable bundles have been used over the years, including the spectral cover construction over elliptically fibered Calabi-Yau three-folds \cite{Friedman:1997yq, Friedman:1997ih, Donagi:1998xe, Andreas:1999ty, Donagi:1999gc, Donagi:1999ez, Donagi:2000zf, Donagi:2000zs, Braun:2005ux, Braun:2005bw, Braun:2005nv, Blumenhagen:2006ux, Blumenhagen:2006wj, Gabella:2008id, Anderson:2019agu}, monad bundles \cite{Distler:1987ee, Kachru:1995em, Anderson:2008ex, Anderson:2008uw, Anderson:2009mh, He:2009wi}, extension bundles \cite{Bouchard:2005ag, Blumenhagen:2006ux, Blumenhagen:2006wj}, as well as direct sums of line bundles \cite{Blumenhagen:2005ga, Blumenhagen:2006ux, Blumenhagen:2006wj, Anderson:2011ns, Anderson:2012yf, Anderson:2013xka, He:2013ofa, Buchbinder:2013dna, Buchbinder:2014qda, Buchbinder:2014sya, Buchbinder:2014qca, Anderson:2014hia, Constantin:2015bea, Buchbinder:2016jqr, Larfors:2020weh}. 

Each of these compactification settings has its own virtues: bundles obtained through the spectral cover construction can be directly used in the study of heterotic/F-theory duality, while monad and extension sequences provide an accessible construction of non-abelian bundles. 
The main virtue of line bundle sums resides in their `split' nature: many of the consistency and phenomenological constraints can be imposed line bundle by line bundle, making this class searchable by systematic methods, at least for manifolds with relatively small Picard number. In this manner, in Ref.~\cite{Anderson:2013xka} an exhaustive search\footnote{While the space of line bundle sums of a fixed rank over a given manifold is unbounded, it was noticed that phenomenologically viable models  correspond to line bundle sums where all entires are relatively small integers, an observation which effectively renders the search space finite, though typically very large (e.g.~the size of the search space involved in Ref.~\cite{Anderson:2013xka} was of order $10^{40}$ bundles). 
} for $SU(5)$ GUT models has been accomplished for Calabi-Yau three-folds with non-trivial fundamental group and a Picard number smaller than~$7$, the search being extended in Ref.~\cite{Constantin:2018xkj} to manifolds of Picard number equal to $7$. 

These searches resulted in the largest dataset of three generation $SU(5)$ GUT models derived from string theory to date, with some $10^7$ explicitly constructed models~\cite{Anderson:2013xka, Constantin:2018xkj}. One of the important empirical lessons of these searches was that, if extended to larger Picard number manifolds, this class of compactifications would produce at least $10^{23}$ and very likely up to $10^{723}$ three generation models \cite{Constantin:2018xkj}. This is certainly good news, as the string phenomenology experience accumulated over the last few decades suggests that it is staggeringly difficult to fine-tune any particular construction to simultaneously meet all the properties of the Standard Model. Having at hand a huge number of good starting points (three generation models) brings about much better prospects. However, in order to cope with the large exponents the systematic scanning approach needs to be replaced with more effective methods of search. 

\subsection{A model builder's to-do list} What lies in front of the model builder is a list of non-trivial steps:
\begin{itemize}[leftmargin=12pt]\itemsep0em
\item[1.] Consider Calabi-Yau threefolds $X$ from existing databases, such as the list of $\sim\!\!8000$ CICYs \cite{Candelas:1987kf, Green:1987cr}, the Kreuzer-Skarke dataset of Calabi-Yau hypersurfaces in four-dimensional toric varieties of around half a billion \cite{Kreuzer:2000xy}, as well as the more recently constructed generalised CICYs~ \cite{Anderson:2015iia} and Gorenstein Calabi-Yau threefolds~\cite{Schenck:2020bok}. Most of these manifolds are simply connected which renders them unusable at Step 2, hence the need to look for discrete, freely acting groups $\Gamma:X\rightarrow X$ in order to construct smooth quotients $ X / \Gamma$ with fundamental group~$\Gamma$. 
To date there are a few hundred known examples of Calabi-Yau threefolds with non-trivial fundamental group~\cite{Candelas:2008wb, Braun:2010vc,Candelas:2010ve, Candelas:2015amz, Candelas:2016fdy, Constantin:2016xlj, Braun:2017juz, Larfors:2020weh}. 
\item[2.] Construct holomorphic stable bundles $V$ over $X$ such that the four-dimensional compactification contains the Standard Model gauge group $SU(3) \times SU(2) \times U(1)$. This step is usually realised in two stages\footnote{In Ref.~\cite{Anderson:2014hia} it was shown that it is not feasible to directly break $E_8$ to the Standard Model group because the large number of conditions that have to be imposed in order to obtain a correct physical spectrum in the absence of an underlying grand unified theory is incompatible with gauge coupling unification. }, by firstly breaking $E_8$ to one of the standard GUT groups and then breaking the latter to the Standard Model gauge group using Wilson lines. Since the final model is constructed on the quotient $X/\Gamma$, the bundle $V$ needs to be $\Gamma$-equivariant in order to descend to a bundle on the quotient threefold, which is non-trivial to check. Checking stability is also a difficult step, in general~\cite{Anderson:2008ex}. The bundle $V$ and the tangent bundle $TX$ also have to satisfy the anomaly cancellation condition.
\item[3.] Derive the matter spectrum of the four-dimensional theory and check that it matches the MSSM spectrum. The fermion fields in the low energy theory correspond to massless modes of the Dirac operator on the internal space, counted by bundle-valued cohomology groups on $X$. This step involves checking: (a) the number of generations, which is relatively easy to compute as a topological index and (b) the presence of a Higgs field and the absence of any exotic matter charged under the Standard Model gauge group, both of which requiring knowledge of cohomology, which can be computationally expensive, in general. Typically, only a small fraction of models have the exact MSSM spectrum. 
\item[4.] Constrain the resulting Lagrangian, in order to avoid well-known problems of supersymmetric GUTs, such as fast proton decay. For this purpose, additional discrete or continuous symmetries derived from the compactification set-up can be essential.
\item[5.] Derive information about the detailed properties of the model, such as holomorphic Yukawa couplings, fermion mass-terms and $\mu$-terms. In a first step, these quantities can be extracted from the holomorphic superpotential of the theory using techniques from algebraic and/or differential geometry.
\item[6.] Compute physical Yukawa couplings. For this, the kinetic terms for the matter fields need to be rendered in canonical form, a computation that requires the explicit knowledge of the Calabi-Yau metric on $X$ and the gauge connection on $V$. Except in very special cases, these quantities are not known analytically and are very hard to obtain numerically. As reviewed below, the recent use of machine learning techniques has significantly improved the efficiency of such computations, making feasible the calculation of physical couplings. 
\item[7.] Stabilise the unconstrained continuous parameters of the internal geometry (moduli fields). Spontaneously break supersymmetry and compute soft supersymmetry-breaking terms.
\end{itemize}

Every phenomenological requirement in this list leads to a substantial reduction in the number of viable models. As such, it is crucial to start with a large number of models or else the chances of retaining a realistic model in the end are extremely limited. Constructing a large number of models by hand is impractical. 
Systematic automated searches have their own limitations, despite substantial advancements in the computational power. Pushing further the boundaries of the explorable part of the string landscape requires a new approach, based on heuristic methods of search such as reinforcement learning and genetic algorithms, to the discussion of which we now turn. 

\section{Reinforcement Learning and Genetic Algorithms}
\subsection{Reinforcement Learning}\label{sec:RL}

Reinforcement learning (RL) is a machine learning approach in which an artificial intelligence agent self-trains to make a sequence of decisions in order to achieve a specified goal within a large and potentially complex environment. The environment corresponds to the space of potential solutions for a given problem. The navigation is aided by a set of rewards and penalties, specified by the programmer, which guide the machine's learning process. 
The RL agent self-trains without any prior knowledge of the environment, a feature that distinguishes RL from supervised and unsupervised learning. 

Every state of the environment, that is every potential solution to the given problem is associated with a numerical value reflecting how well it fits the properties sought from target solutions. 
The learning process relies on this {\itshape intrinsic value function}, and different such functions can lead to very different kinds of performance. 
The search is then divided into multiple {\itshape episodes} involving a fixed number of maximal states. Typically, the initial state  is randomly chosen and the episode ends either when a target state is found or when the maximal episode length is reached. The progression of states within an episode is dictated by the current {\itshape policy}, which is initially a random function that gets corrected using a neural network after each episode or periodically after a fixed number of episodes specified by the programmer during the self-training phase. Training is usually stopped once the agent is capable of reaching a target state from virtually any starting point. The typical maximal length of the episodes can be estimated in the following way. If the space of solutions is a $d$-dimensional hypercube of length~$l$, then 
\begin{equation}
\text{typical maximal episode length }\sim d^{1/2}l~,
\end{equation}
which is the length of the diagonal, the idea being that within an episode the agent should have enough `time' to travel between any two points of the search space. A longer episode length gives the agent more `time' to find a good solution within any given episode, however it can also determine it to become fixated on a small number of terminal states. For this reason it is customary to introduce a penalty on the episode length, giving the agent an incentive to find terminal states that are as close as possible to the original random starting point. On the other hand, if the search space contains sizeable `gaps' with no target states, the episode length should be large enough so that the agent can move out of these regions within an episode. Often the distribution of target states in the search space is not known, which makes the episode length an important hyper-parameter that needs adjustment.
\begin{figure}[ht]
\centering
\includegraphics[width=0.24\textwidth]{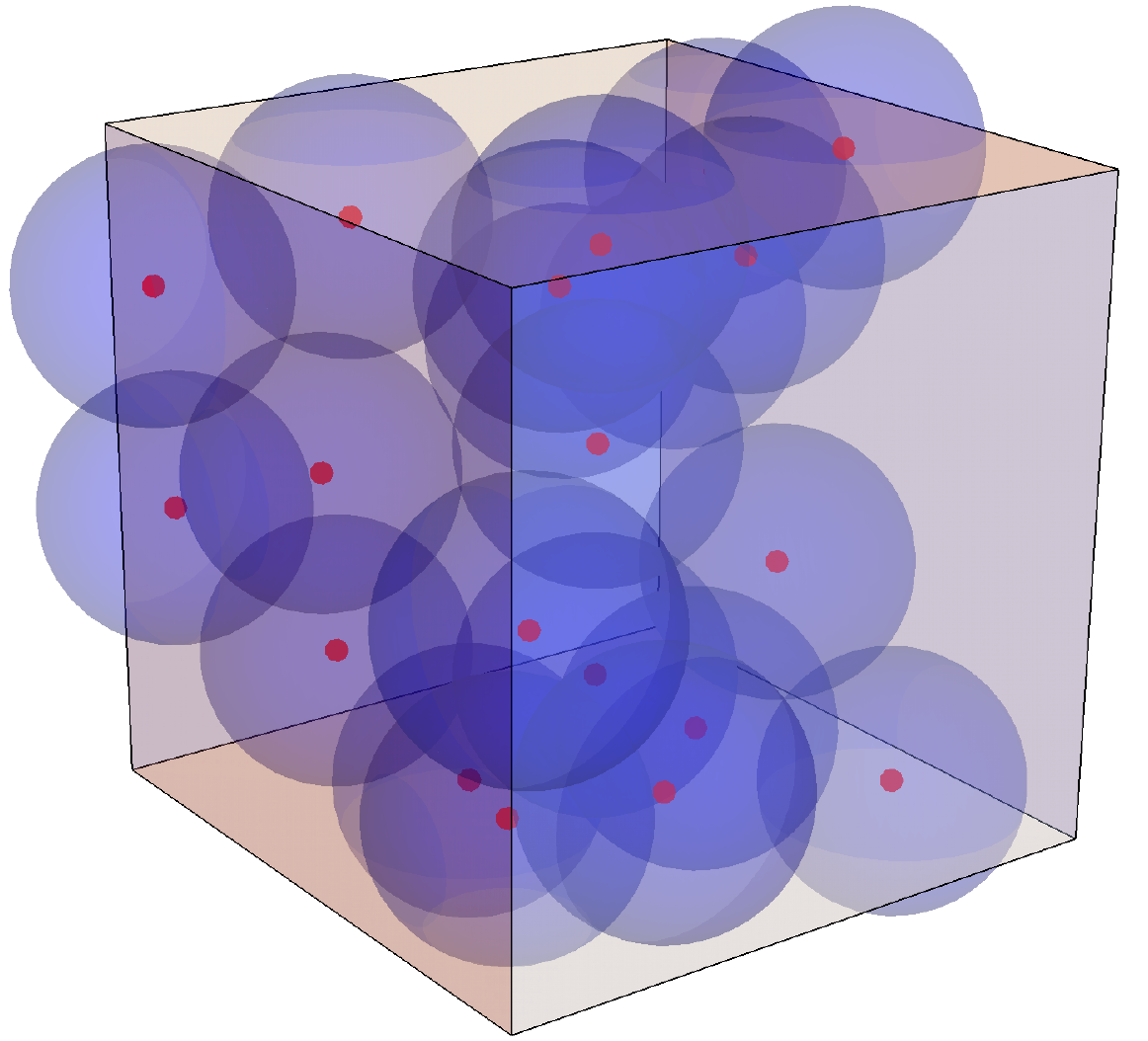}
\caption{An idealised picture of the search space and target states.}
\label{Fig:SearchSpace}
\end{figure}

Ideally, after sufficiently many training episodes, the AI agent `knows' enough about the landscape (the intrinsic value function) to (1) reach a terminal state for virtually any initial random point and (2) reach any specific target state within an episode provided that the initial random point is close enough. For the purpose of illustration, Figure~\ref{Fig:SearchSpace} shows a situation where the target states (the red points) are uniformly distributed. If the initial random point of an episode falls within a blue ball (basin of attraction), the corresponding target state will be found within that episode. The basins of attraction cover the entire space, which means that the number of episodes needed to obtain all the target states is, in principle, comparable to the total number of target states available in the search space. Of course, in practice the basins of attraction are not spheres; they roughly correspond to level hypersurfaces of the value function, though their exact shape depends very much on the training history. However, the main idea remains essentially the same: the target states get `thickened', acting as attractor points within the corresponding attractor basins.

\subsection{Fixed and dynamical elements of RL}
We review here some of the basic ideas and terminology used in RL referring the reader to Refs.~\cite{sutton2018reinforcement, Ruehle:2020jrk} for more in-depth accounts. We divide the presentation into fixed and dynamical elements. The fixed (hard-wired) elements include: 
\vspace{-8pt}
\begin{itemize}\setlength\itemsep{0em}
\item[{\itshape i.}] The {\itshape environment}, consisting of a set $\mathcal S$ of {\itshape states}, typically of very large size. Often the states are represented as numerical lists (vectors). At every instance the agent (the computer programme) is in one of the states. The agent performs a sequence of steps, which leads to a notion of time. 
\item[{\itshape ii.}] The set of possible actions $\mathcal A(s)$ that can be taken from a state $s\in \mathcal S$ to move to other states. In the simplest situation this set is the same for all states $s$ and denoted by $\mathcal A$.
\item[{\itshape iii.}] At time step $t$ the agent moves from state $s_t$ to a new state $s_{t+1}$ by performing an action~$a_t$. The choice of action is dictated (deterministically or probabilistically) by a {\itshape policy}. The policy is essentially dynamical, but its initialisation is fixed by the programmer. Often the initial policy consists of a random choice of actions at every step. 
\item[{\itshape iv.}] A characterisation of {\itshape terminal states}, defining a subset $\mathcal T\subset \mathcal S$. Not every environment has terminal states, but when it has, these are the targets towards which the agent moves. 
\item[{\itshape v.}] The maximal length of an {\itshape episode}, $l_{\rm ep}$. The search is divided into episodes $\mathcal E_i\subset \mathcal S$ which end either when a terminal state is reached or after a certain number of steps specified by the maximal length. 
\item[{\itshape vi.}] The method of {\itshape sampling} the initial states of episodes. This can be completely random or specified by a probability distribution. 
\item[{\itshape vii.}] The {\itshape reward function}. This is the essential tool in controlling the agent since the system of rewards and penalties is the only way to communicate with the learning process and to modify the existing policy. Typically the reward is constructed as a real number $r(s,s')$ associated with each possible step $s\rightarrow s'$. Thus at time step $t$ the agent receives a reward $r_t$ associated with $s_t\rightarrow s_{t+1}$. 
\item[{\itshape viii.}] Often, it is appropriate to construct the reward function by first defining the {\itshape intrinsic value of states}, $V:\mathcal S\rightarrow \mathbb R$ as a measure of how badly a state $s$ fails to achieve the properties expected from a terminal state. For instance, if $V$ is semi-negative, the reward $r(s,s')$ can be chosen as $V(s')-V(s)$, giving an incentive to move towards states of higher value. To this reward function one can add, e.g.~a fixed penalty for each step, a penalty for stepping outside of the environment, as well as a typically large bonus for reaching a terminal state. 
\item[{\itshape ix.}] In general, it is not wise to make judgements (update the policy) based on immediate rewards alone. In order to take into account delayed rewards, one defines a {\itshape return function} $G:\mathcal S\rightarrow \mathbb R$ as a weighted sum 
\begin{equation}
G_t = \sum_{k\geq 0} \gamma^k r_{t+k}~,
\end{equation} 
computed for a state $s_t$ by following the trajectory dictated (deterministically or probabilistically) by the policy and adding the weighted rewards until the end of an episode. Of course, the numbers $\{G_t\}$ associated with the states $\{s_t\in\mathcal E_i\}$ can only be computed once the episode $\mathcal E_i$ ends. The number $\gamma\in [0,1)$ is called {\itshape discount factor}. It is sub-unitary in order to give a greater weight to rewards in the immediate future, and typically very close to $1$ in order to give some reasonable weight to rewards arising in the more distant future. 
\item[{\itshape x.}] The size of the {\itshape batch}. The agent collects data in the form of triplets $(s_t, a_t, G_t)$ and communicates it in batches to the algorithm controlling the policy. The policy is then updated and a new batch is collected and communicated back. Thus the learning process follows an iterative approach. 

The learning curve is typically very sensitive to the architecture and some scanning is required to fix hyperparameters such as the maximal length of episodes $l_{\rm ep}$, the discount factor~$\gamma$ and the batch size.
\end{itemize}

Learned (dynamical) elements. There are various flavours of RL and they differ in the quantities that are being learnt. In general, there are three learnt elements: 
 
\vspace{-8pt}
\begin{itemize}\setlength\itemsep{0em}
\item[{\itshape i.}] The {\itshape policy}. In deterministic approaches the policy is a map $\pi:\mathcal S\rightarrow \cA$ specifying what action $a=\pi(s)$ needs to be taken if state $s$ is reached. In stochastic approaches, the policy is a probability distribution $\pi:\cA\times \mathcal S\rightarrow [0,1]$, specifying the probability $\pi(a|s)$ for action $a$ to be picked in state $s$. 
The aim of the game is to find a policy that brings a maximal return (in the deterministic case) or a maximal expected return (in the stochastic case). RL provides a method of obtaining an approximately optimal policy in this sense. The method is not guaranteed to work in all cases and the algorithm may often lead to policies that are only locally optimal, as they essentially rely on a local search. As already mentioned, even in the cases where the method eventually proves to be successful, a certain amount of fine tuning of the hyperparameters is usually required. 
In the stochastic setting described below the policy is a function taking as input a state of the environment, represented by a numerical list, and outputting a list of probabilities, one for each possible action. 
\item[{\itshape ii.}] The {\itshape state-value function} under policy $\pi$ is defined as the expected return  when starting in state  $s$ and following the policy $\pi$ thereafter, $v_\pi(s) = \mathbb E_\pi[G_t | s_t = s]$. The value of a terminal state is zero, since there are no future returns in this case. 
\item[{\itshape iii.}] The {\itshape action-value function} under policy $\pi$ is the expected return starting from state $s$, taking the action $a$, and thereafter following policy $\pi$,  $q_\pi(s,a) = \mathbb E_\pi[G_t | s_t = s, a_t = a]$. The value of a state $v_\pi(s)$ depends on the values of the actions possible in that state and on how likely each action is to be taken under the current policy.  
\end{itemize}

In the simplest setting, called REINFORCE, the aim is to learn a parameterised policy that can select actions without consulting a value function. 
A brute force optimisation would run over all possible policies, sample returns while following them and then choose the policy with the largest expected return. Unfortunately, the number of possible policies is typically very large or infinite, making brute force optimisation unfeasible. 

For this reason, in RL algorithms the policy $\pi$ is controlled by a neural network with internal parameters (weights and biases) collectively denoted by $\boldsymbol{\theta}$. The network takes as input a state $s\in\mathcal S$ and outputs a list of probabilities, denoted by $f_{\boldsymbol{\theta}}(s)$. Then, if the $a$-th action is represented by the $a$-th unit vector in $\mathbb R^{|\cA|}$, the probability to choose action $a$ in state $s$ will be the dot product $\pi_{\bf \boldsymbol{\theta}}(a|s)=a\cdot f_{\boldsymbol{\theta}} (s)$. The internal parameters $\boldsymbol{\theta}$ get corrected after the analysis of each batch, so as to minimise the loss function defined below. REINFORCE uses triplets of data $(s_t,a_t,G_t)$ which include the complete return $G_t$ from time $t$, that is all future rewards up until the end of the episode. The internal parameters $\boldsymbol{\theta}$ get updated after the analysis of the $(s_t,a_t,G_t)$ data triplet in the following way (see Ch.~13 of Ref.~\cite{sutton2018reinforcement}): 
\begin{equation}
\boldsymbol{\theta}_{t+1} = \boldsymbol{\theta}_t + \alpha\, G_t \frac{\nabla \pi_{\boldsymbol{\theta}_t}(a_t,s_t)}{\pi_{\boldsymbol{\theta}_t}(a_t,s_t)}~,
\end{equation}
where $\alpha$ is the learning rate specified by the programmer as a hyperparameter. Put differently, the neural network is trained on the loss function $L(\boldsymbol{\theta}_t)= G_t\ln(a_t\cdot f_{\boldsymbol{\theta}_t}(s_t))$.

\vspace{4pt}	
There are many other flavours of RL. For instance, a version of RL called actor-critic introduces, apart from the policy network discussed above, a second network which controls the state-value function. The agent continues to follow $\pi$, but the performance of the policy is judged by the second network. Both networks are used to improve each other in this case. %

\subsection{Genetic algorithms}
Genetic algorithms are a class of heuristic problem solving methods inspired by evolutionary biology. The idea is to encode the data specifying a solution attempt into a sequence of (binary) digits. A population of such solution attempts is created and evolved according to a fitness function which gives a measure for how close the attempt is to an actual solution. The fitness function corresponds to what we called intrinsic value function in the context of RL.  

The optimal size of the population, $N$, depends logarithmically on the length of the sequence of digits encoding solution attempts. Typically $N$ is of order of a few hundred individuals. The initial population can be generated randomly or seeded around areas of the solution space where optimal solutions are likely to be found. 

Evolution then proceeds by selection, breeding and mutation. 
A popular choice for the {\itshape selection} method is to start by ranking the solution attempts according to their fitness. An individual at rank $k$ is then selected for breeding with a probability that depends linearly on its ranking, such that the probability for the top individual $P_1$ is equal to a multiple $\alpha$ of the probability $P_N$ of the least fit individual. Typically, $\alpha$ is chosen in the range $2\leq\alpha\leq 5$. While the fittest individuals have a higher chance to reproduce, the scheme also ensures that the less fit individuals are also able to breed, which preserves a healthy variety of `genes' throughout the evolutionary process. 
	
The {\itshape breeding} is usually implemented as an $M$-point cross-over by which the two binary sequences are cut at the same $M$ random points and the cut sections are alternatively swapped. This implementation is made possible by the fact that in the simplest setting all binary sequences have a fixed length. There exist other, more sophisticated versions of genetic algorithms, including genetic programming, where solution attempts are represented as bit strings of variable size, or as trees/graphs. In these cases the cross-over implementation is more complex. 
Often a single point cross over turns out to perform well enough. 
Once a new generation is formed through cross-over, a small fraction (usually around one percent) of the binary digits, selected randomly, are flipped. These {\itshape mutations} ensure that the population does not stagnate and continues to evolve towards better solutions or towards different optimal solutions. For the applications envisaged here, the optimal solution is not unique; rather there are many optimal solutions sparsely scattered over a huge landscape. 
Finally, one can invoke an element of {\itshape elitism}: in order to ensure that the new generation has a greater or equal maximum fitness than the previous generation, the fittest individual(s) from the previous generation can be copied into the new one replacing the least fit next individual. 

The process is then repeated over many generations, and terminates after a pre-defined number of evolutionary cycles $N_{\rm gen}$. This number can be chosen by trial and error. It needs to be large enough to allow the algorithm to find a sufficiently large number of optimal solutions. On the other hand, the typical situation is that beyond a certain number of cycles very few new solutions are found, indicating that the search can stop and restart from a different random initialisation.

Compared to Reinforcement Learning, Genetic Algorithms benefit from the advantage of a simpler implementation as well as from the absence of a training phase. On the other hand, the success or failure of each method very much depends on the problem in question, so having available several complementary methods can be crucial in tackling certain problems. For situations where both methods turn out to be successful, they can be used in conjunction in order to estimate the achieved degree of comprehensiveness in finding most of the solutions present in the environment.

In particle physics and string theory genetic algorithms have not yet been widely used. The first application in string theory was undertaken in Ref.~\cite{Abel:2014xta} for heterotic model building in the Free Fermionic formulation. More recently, RL has been used in Ref.~\cite{Halverson:2019tkf, Loges:2021hvn} to generate type~IIA intersecting brane configurations that lead to standard-like models and in Refs.~\cite{Larfors:2020ugo, Constantin:2021for} to construct $SU(5)$ and $SO(10)$ string GUT models. The landscape of type IIB flux vacua was  explored in Ref.~\cite{Cole:2019enn} using GAs and Markov chain Monte Carlo methods, while in Refs.~\cite{Krippendorf:2021uxu, Cole:2021nnt} the same methods were used, as well as RL. Other applications of RL include the construction of quark mass models~\cite{Harvey:2021oue}, solving the conformal bootstrap equations~\cite{Kantor:2021jpz} and learning to unknot~\cite{Gukov:2020qaj}. 

It is important to note that RL and GAs are qualitatively different from the more standard supervised and unsupervised learning techniques, which have also been recently used in the exploration of the heterotic string landscape~\cite{Parr:2019bta, Mutter:2018sra, Faraggi:2019iic, Deen:2020dlf}.

\section{Model building with monad bundles and line bundle sums}
In this section we look at the details of heterotic model building on smooth Calabi-Yau threefolds with holomorphic bundles constructed either as monad bundles or as sums of line bundles. These classes of compactifications have proven to include many phenomenologically attractive models~\cite{Anderson:2008ex, Anderson:2008uw, Anderson:2009mh, He:2009wi, Blumenhagen:2005ga, Blumenhagen:2006ux, Blumenhagen:2006wj, Anderson:2011ns, Anderson:2012yf, Anderson:2013xka, He:2013ofa, Buchbinder:2013dna, Buchbinder:2014qda, Buchbinder:2014sya, Buchbinder:2014qca, Anderson:2014hia, Constantin:2015bea, Buchbinder:2016jqr, Braun:2017feb, Braun:2018ovc, Otsuka:2018oyf, Otsuka:2018rki, Larfors:2020weh}, hence the motivation to explore them further. Our aim here will be to understand the extent to which the heuristic search methods discussed above can speed up the search for realistic models. 

The discussion at the end of Section~\ref{sec:RL} suggested that after the initial self-training stage, RL networks have the capacity to guide the search towards a terminal state from virtually any starting point in the environment within a small number of steps, thus splitting the environment into basins of attractions. Finding all terminal states then amounts to finding one starting point in each basin of attraction. If basins of attraction are of roughly the same size, the computational time required to find most of the solutions scales linearly with the number of terminal states present in the environment. This has to be contrasted with the case of systematic scans where the computational time scales linearly with the total number of states contained in the environment. 

The same (and, in fact a better) behaviour in terms of computational time has been observed in the case of GAs~\cite{Abel:2021rrj, Abel:2021ddu}. 
As such, RL and GAs can be extremely efficient in exploring spaces that are too large and the desirable states too sparse to be found by systematic scans or by random searches. More interestingly, and somewhat counterintuitively, the fact that the computational time required to find most of the solutions scales with the number of optimal solutions, rather than the total size of the environment, implies that the search is more efficient when more constraints are being included. In any systematic scan including more constraints comes with an inevitable computational cost, however with RL or GAs this cost can be overcompensated by the (typically substantial) reduction in the number of basins of attraction corresponding to different solutions that satisfy all the constraints.  Our focus in the following sections will fall on identifying which constraints can be currently implemented in automated searches or are susceptible of implementation in the near future given certain theoretical advancements, such as the discovery of explicit analytic formulae for cohomology dimensions.

For string theory model building, the reduction in computational time can be a real game changer. The size of the search spaces is typically very large. For a fixed Calabi-Yau threefold~$X$, infinite classes of topologically distinct bundles can be considered. However, the experience of various systematic scans indicates that viable models can only be found in a finite search region whose size scales exponentially like $10^{\alpha h^{1,1}(X)}$ with a multiple $\alpha$ of the Picard number $h^{1,1}(X)$. The number $\alpha$ depends on the details of the class of bundles in question but is generally greater or equal to~$5$. The size of the solution space, on the other hand, is much smaller. To give an estimate figure, we refer to the comprehensive study of line bundle sums leading to $SU(5)$ models with three families undertaken in Refs.~\cite{Anderson:2013xka, Constantin:2018xkj}, which found a number of $10^{h^{1,1}(X)}$ solutions for a typical Calabi-Yau threefold~$X$. The change in the exponent is significant and by adding more physical constraints the size of the solution space is bound to decrease further.

\subsection{Monad bundles} 
A monad bundle $V$ on a complex manifold $X$ is constructed from two sums of holomorphic line bundles $B$ and $C$, via the short exact sequence
	\begin{equation}\label{eqnMonadSeqenceIntro}
		0 \rightarrow V \rightarrow B \stackrel{f}{\rightarrow} C \rightarrow 0, 
	\end{equation}
where $f$ is a bundle morphism and, by exactness, $V={\rm ker} (f)$. Each line bundle in $B$ and~$C$ is specified by its first Chern class, which in a basis of the second cohomology of $X$ corresponds to a list of $h^{1,1}(X)$ integers. This means that $V$ is specified by $h^{1,1}(X)\left( {\rm rk}(B)+{\rm rk}(C) \right)$ integers. $V$ also depends on the monad map~$f$, which encodes the bundle moduli. The map $f$ can be assumed to be generic as long as the rank of $V$, given by the dimension of ${\rm ker}(f)$, is constant across $X$. This condition guarantees that $V$ is a bundle rather than a more general type of sheaf. When $V$ is a bundle its rank is given by ${\rm rk}(V)={\rm rk}(B)-{\rm rk}(C)$. 

Provided that $f$ is generic enough, the structure group of $V$ is $U({\rm rk}(V))$. In order to break the heterotic $E_8$ gauge symmetry to one of the standard GUT symmetry groups, $SU(5)$, $SO(10)$ or $E_6$, the structure group of $V$ has to be of special type, which implies that $c_1(V)=0$, and hence $c_1(B) =c_1(C)$. This condition reduced the number of integers specifying $V$ to $h^{1,1}(X)\left( {\rm rk}(B)+{\rm rk}(C)-1 \right)$. 

Since the line bundle integers specifying $B$ and $C$ can take arbitrary values, the search space is infinite. To make it finite, one can allow these integers to run in a finite range, for instance between $-4$ and $5$, which turns out to be the range where most of the good models lie. In this case, the size of the search space is of order
$10^{h^{1,1}(X)\left( {\rm rk}(B)+{\rm rk}(C) -1 \right)}$.
Since ${\rm rk}(C) \geq 1$ and ${\rm rk}(V)=4$ for $SO(10)$ models, while for $SU(5)$ models ${\rm rk}(V)=5$, it follows that the size of the search space is 
\begin{equation}
10^{\geq 5\,h^{1,1}(X)}~.
\end{equation}
The computational time required to perform even the most basic checks for a monad bundle being of order of a few mili-seconds on a standard machine, this implies that for any manifold with $h^{1,1}(X)\geq 2$ a systematic and comprehensive search is not possible (or just about possible in the case $h^{1,1}(X)=2$).

In Refs.~\cite{Constantin:2021for, Abel:2021rrj, Abel:2021ddu} it was shown that, despite its gigantic size, this class of heterotic string compactifications is searchable by means of RL and GA methods. More specifically, the studies concentrated on monad bundles leading to $SO(10)$ supersymmetric GUT models. For group-theoretical reasons, the Wilson-line breaking of $SO(10)$ to the Standard Model requires a discrete group $\Gamma$ which is at least $\mathbb{Z}_3\times \mathbb{Z}_3$. Unfortunately there are not many known Calabi-Yau threefolds admitting a freely acting symmetry group of this size~\cite{Braun:2010vc,Braun:2017juz, Candelas:2016fdy, Constantin:2021for}, so the choice of manifold in this case is rather constrained. As such, these studies  focused on a few manifolds realised as complete intersections in products of projective spaces. The simplest of these  is the bicubic threefold represented by the configuration matrix
\begin{equation}\label{eq:CYs}
X= \left[\begin{array}{c|c}\mathbb{P}^2&3\\\mathbb{P}^2&3\end{array}\right]^{2,83}_{-162}\;.
\end{equation}
It is worth noting that the search space has a large degeneracy. For the bicubic, equivalent bundles arise from permuting the two $\IP^2$-factors in the embedding, as well as from permuting the line bundles in $B$ and $C$. This amounts to a group of order $2!\cdot 6!\cdot2! = 2800$. 
\vspace{8pt}

Both the RL and GA implementations turned out to be successful in identifying models, termed `perfect states',  that pass the following checks:
\begin{itemize}\itemsep0em
\item[a)] a sufficient criterion for checking the bundleness of $V$;
\item[b)] the anomaly cancellation condition;
\item[c)] the Euler characteristic being equal to $-3|\Gamma|$, where $\Gamma$ is a freely acting symmetry on $X$;
\item[d)] a necessary condition for the equivariance of $V$ with respect to the symmetry $\Gamma$;
\item[e)] a number of necessary conditions for the stability of $V$ relying on Hoppe's criterion and the availability of explicit line bundle cohomology formulae on $X$.
\end{itemize}

Including more checks in the search algorithm would require further theoretical progress. For instance, for properties such as the full spectrum or bundle stability analytical formulae for cohomology dimensions of monad bundles would be a crucial ingredient. 

\begin{figure}[ht]
     \centering
    {\includegraphics[width=0.59\textwidth]{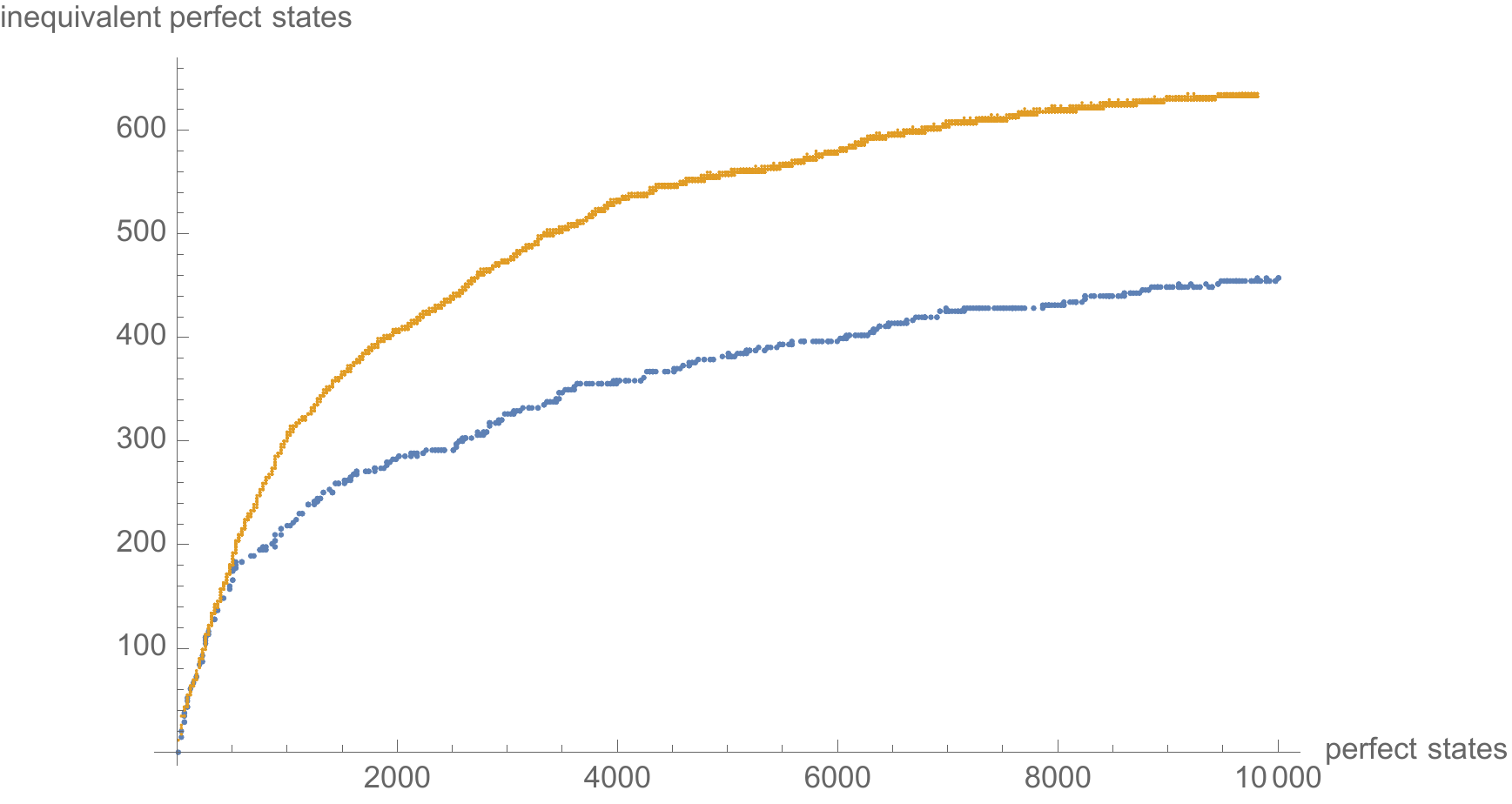}}
       \caption{Saturation of the number of inequivalent `perfect models'. The orange curve corresponds to RL search and the blue curve to the GA search. The RL search took $35$ core days, while the GA search took only $1$ core day. A total of $\sim\!700$ inequivalent models have been found, with an overlap of over $90\%$ between the two methods.}
     \label{fig:BicubicRLSaturation}
\end{figure}

The RL/GA explorations of monad bundles on the bicubic manifold accomplished a high degree of comprehensiveness in finding all the models satisfying the above criteria using relatively modest computational resources. Indirect evidence in support of this claim was obtained by exploiting the degeneracy of the environment. As shown in Figure~\ref{fig:BicubicRLSaturation}, the number of inequivalent perfect models found in the search saturates as a function of the total number of perfect models found, suggesting that most of the inequivalent perfect models have been found. Moreover, comparing the results of the GA search with those obtained through RL, it turns out that the two datasets of inequivalent models have an overlap of over $90\%$, despite the great differences distinguishing the two methods. 
This also suggests that the details of the optimisation process are not really essential once the processes begin to saturate, provided that they share the same incentives. 

Comparing the efficiencies of the two methods, it turns out that on this environment GA is, overall, more efficient by about an order of magnitude than RL in identifying models that pass all the above criteria. In general, such a comparison would be difficult to make due to the intrinsic differences between the two methods, however in this case the comparison is legitimate as it refers to the time taken to accomplish a sufficiently high degree of comprehensiveness.

The methods described above can be easily extended to other contexts, for instance to rank~$5$ monad bundles leading to $SU(5)$ models. In this case there are no group-theoretic restrictions on the freely acting symmetry $\Gamma$ (e.g.~$\Gamma$ can be as small as $\mathbb Z_2$), so many more choices for $X$ are available. As such, the expectation is that vastly larger numbers of $SU(5)$ models can be found using RL, GAs and monad bundles.

\subsection{Line bundle sums} 
 There are several  advantages to working with sums of line bundles, as opposed to irreducible vector bundles. Firstly, such configurations are relatively simple to deal with from a computational point of view. Secondly, Abelian models are characterised by the presence of  additional $U(1)$ gauge symmetries, which are broken at a high-energy scale, but remain in the low-energy theory as global symmetries, constraining the resulting Lagrangian and giving more information than is usually available in other constructions.  Finally, although line bundle sums represent special loci in the moduli space of vector bundles of a given topology, these simple configurations provide a computationally accessible window into a bigger moduli space of heterotic compactifications: if a line bundle sum corresponds to a standard-like model, then usually it can be deformed into non-Abelian bundles that also lead to standard-like models \cite{Buchbinder:2013dna, Buchbinder:2014qda}. Moreover, the effect of the $U(1)$ symmetries persists even beyond the locus where the bundle splits into a direct sum~\cite{Buchbinder:2014sya}.

In Ref.~\cite{Anderson:2013xka} a data set of about $10^6$ models with $SU(5)$ gauge group and the correct chiral asymmetry has been constructed on CICYs with Picard number~$<7$. In Ref.~\cite{Constantin:2018xkj}, this data set expanded by an order of magnitude by considering CICYs with Picard number~$7$. 
It is expected that a significant number of these models will descend to standard-like models after dividing by the corresponding discrete symmetry. However, the detailed analysis of this wealth of models has so far been hindered by the difficulty of (equivariant) cohomology computations.

Cohomology computations indeed represent the main limiting factor in the analysis of heterotic compactifications on Calabi-Yau manifolds with holomorphic vector bundles. For instance, in the search algorithm for monad bundles described above it was not possible to include criteria related to the full description of the low-energy spectrum, as these would rely on slow -- if achievable at all  -- cohomology computations. The situation is different for line bundle sums, due to the recent discovery of analytic formulae for cohomology~\cite{Constantin:2018otr, Constantin:2018hvl, Klaewer:2018sfl, Larfors:2019sie, Brodie:2019ozt, Brodie:2019pnz, Brodie:2019dfx, Brodie:2020wkd, Brodie:2020fiq}. 
The new cohomology formulae allow for a quick check of the entire low-energy spectrum, which represents a significant improvement from the usual check on the number of chiral families, computed as a topological index.  
Concretely, cohomology constraints can be imposed to ensure: 
\begin{itemize}\itemsep0em
\item[a)] three families of quarks and leptons;
\item[b)] the presence of a Higgs field and the absence of any exotic matter charged under the Standard Model gauge group;
\item[c)] a hierarchy of Yukawa couplings compatible with a heavy third generation;
\item[d)] the absence of operators inducing fast proton decay and R-parity violating operators;
\item[e)] the presence of a $\mu$-term and the existence of right-handed neutrinos;
\item[f)] Yukawa unification, etc. 
\end{itemize}
These constraints can be imposed along with the usual set of requirements: 
\begin{itemize}\itemsep0em
\item[g)] the anomaly cancellation condition;
\item[h)] poly-stability of the line bundle sum, that is checking the existence of a non-empty locus in K\"ahler moduli space where the slopes of all line bundles simultaneously vanish;
\item[i)] equivariance with respect to the freely acting discrete symmetry.
\end{itemize}

We illustrate the implementation of the new constraints relying on line bundle cohomology formulae with the discussion of  dimension four proton decay operators in $SU(5)$ GUT models. These operators are of the form  $\bar{\bf 5}\,\bar{\bf 5}\,{\bf 10}$, possibly with a number of singlet insertions, ${\bf 1}{\bf 1}\ldots{\bf 1}\bar{\bf 5}\,\bar{\bf 5}\,{\bf 10}$. In heterotic line bundle models the $SU(5)$ multiplets come with additional $U(1)$ charges, in fact with $S\left( U(1)^5\right)$ charges. These can be represented by vectors $\mathbf{q} = \left( q_1,\ldots, q_5\right)$. The group $S\left( U(1)^5\right)$ consists of elements $\left( e^{i\theta_1},\ldots, e^{i\theta_5}\right)$, such that the sum of the phases $\theta_1+\ldots+\theta_5=0$. Due to this determinant condition, two $S\left( U(1)^5\right)$ representations, labelled by $\mathbf{q}$ and $\mathbf{q'}$ have to be identified, if $\mathbf{q}-\mathbf{q'} \in \mathbb Z\mathbf{n}$, where $\mathbf{n} = \left(1,1,1,1,1\right)$. 
By working out the necessary branching rules, it turns out that each of the ${\bf 10}$-multiplets is charged under a single $U(1)$. Thus, denoting by $\{\mathbf{e}_a\}_{a=1,\ldots,5}$ the standard basis in five dimensions, a multiplet ${\bf 10}$ charged under the $a$-th $U(1)$ can be denoted by ${\bf 10}_{\mathbf{e}_a}$. Similarly, the patterns of charge assignments for the other $SU(5)$ multiplets are ${\bf 1}_{{\bf e}_a - {\bf e}_b}$, ${\bf 5}_{-{\bf e}_a -{\bf e}_b}$, ${\bf \overline{5}}_{{\bf e}_a+{\bf e}_b}$, ${\bf  \overline{10}}_{-{\bf e}_a}$. In any concrete model, that is for any specific sum of five line bundles over the Calabi-Yau manifold, determining the number of $SU(5)$ multiplets of each type and their charge assignments amounts to the computation of line bundle cohomology dimensions.  

In the case of dimension four operators of the form $\bar{\bf 5}_{a,b}\,\bar{\bf 5}_{c,d}\,{\bf 10}_{e}$, the $S\left( U(1)^5\right)$ charge is $\mathbf{e}_a+\mathbf{e}_b+\mathbf{e}_c+\mathbf{e}_d+\mathbf{e}_e$. In order for such operators to be allowed, $a, b, c, d$ and $e$ must all be different. Thus a sufficient constraint for the absence of such operators is that for any triplet of $SU(5)$ multiplets present in the spectrum $(\bar{\bf 5}_{a,b}, \bar{\bf 5}_{c,d},{\bf 10}_{e})$ the values of $a, b, c, d$ and $e$ must have some overlap. This is a combinatorial problem which can be easily decided provided that the full details of the spectrum (i.e.~multiplets and $U(1)$-charges) are known, information which can be quickly gathered during the search using the analytic formulae for cohomology discussed in the following section.
The large number of constraints listed above is expected to select a relatively small number of string compactifications that can accommodate the Standard Model.

\section{Line bundle cohomology formulae}

The standard methods for computing cohomology include algorithmic methods, based on \v{C}ech cohomology and spectral sequences~\cite{cicypackage, cohomCalg:Implementation, cicytoolkit}. However, these methods are computationally intensive and provide little insight into the origin of the results. The effect is that model building efforts are typically limited to trial-and-error searches. The existence of simple analytic formulae for line bundle cohomology can dramatically change this situation.
Initially these formulae were discovered empirically, through a combination of direct observation~\cite{Constantin:2018otr, Constantin:2018hvl, Larfors:2019sie} and machine learning techniques~\cite{Klaewer:2018sfl, Brodie:2019dfx}. 

The main observation was that line bundle cohomology dimensions on many manifolds of interest in string theory, of complex dimensions two and three, appear to be described by formulae which are essentially piecewise polynomial. 
More precisely, the Picard group decomposes into a (possibly infinite) number of polyhedral chambers, in each of which the cohomology dimensions are captured by a closed form expression. This observation holds for both the zeroth cohomology and the higher cohomologies, with a different chamber structure emerging in each case.  
The mathematical origin of these formulae has been uncovered in Refs.~\cite{Brodie:2019ozt, Brodie:2020wkd} for the case of complex surfaces,  and partially uncovered in Ref.~\cite{Brodie:2020fiq} for Calabi-Yau threefolds (see Ref.~\cite{Brodie:2021zqq} for a recent review).

\subsection{Algebraic results}
For complex surfaces it suffices to understand the zeroth cohomology function $h^0(X,V)$. The formulae for the first and second cohomologies then follow by Serre duality and the Atiyah-Singer index theorem. The zeroth cohomology formulae can be traced back to: (i)~the existence of a fundamental region (usually the nef cone) in which the zeroth cohomology can be equated to the Euler characteristic due to the vanishing of all higher cohomologies and (ii) the existence of a projection map, constructed using Zariski decomposition, that preserves the zeroth cohomology and relates line bundles from the outside of the fundamental region to line bundles inside this region. In Ref.~\cite{Brodie:2020wkd} it was shown that the nef cone data and the Mori cone data are sufficient to determine all line bundle cohomologies on several classes of complex surfaces, including compact toric surfaces, weak Fano surfaces (generalised del Pezzo surfaces), and K3 surfaces. 

For Calabi-Yau three-folds qualitatively new phenomena arise. In Ref.~\cite{Brodie:2020fiq} some of the mathematical structures underlying the empirical formulae for the zeroth line bundle cohomology dimensions on Calabi-Yau threefolds were identified. In particular, it was shown that the zeroth line bundle cohomology encodes a wealth of information about the flops connecting the birational models of the manifold, as well as about Gromov-Witten (GW) invariants. 
It was also noticed that the effective cone (containing all the line bundles with global sections, i.e.~with a non-trivial zeroth cohomology group) decomposes into cohomology chambers where the zeroth cohomology can be expressed as a topological index. The chambers were understood to be either (i) K\"ahler cones of birational models of $X$, inside which the zeroth cohomology can be equated to the Euler characteristic computed on the flopped manifold, or (ii) Zariski chambers, analogous to those arising in the two-dimensional case, where a cohomology-preserving projection operates. 
Moreover, it was understood that the vast majority of known Calabi-Yau threefolds admit flops, many of them flopping to manifolds isomorphic to themselves. In particular, many threefolds admit infinite sequences of flops (and hence an infinite number of zeroth cohomology chambers) and have an infinite number of contractible rational curves~\cite{Brodie:2021toe, Brodie:2021ain, Brodie:2021nit}.
\vspace{8pt}

We illustrate the discussion of cohomology formulae with an example that has been previously studied in Refs.~\cite{Constantin:2018otr,Buchbinder:2013dna, Constantin:2018hvl}. However, note that the earlier formulae were incomplete as they did not take into account the infinite number of chambers that arise in zeroth cohomology. Consider a generic hypersurface of multi-degree $(2,2,2,2)$ in $(\mathbb P^1)^{\times 4}$, corresponding to a smooth Calabi-Yau threefold $X$, known as the tetra-quadric. The manifold has $h^{1,1}(X)=4$ and $h^{2,1}(X) = 68$. The K\"ahler cone $\mathcal K(X)$ descends from the K\"ahler cone of $(\mathbb P^1)^{\times 4}$ and we denote by $\{J_i\}_{i=1,\ldots,4}$ its generators, which are the pullbacks to $X$ of the four $\mathbb P^1$ K\"ahler forms. A line bundle $L$ over $X$ is then specified by its first Chern class $c_1(L) = \sum_{i=1}^4k_iJ_i$, where $k_i$ are integers. The Euler characteristic of $L$ is
\begin{equation}\label{eq:7862_index}
\chi(X,L) = \int_X {\rm ch}(L)\cdot {\rm td}(X) = 2 (k_1 + k_2 + k_3 + k_4 + k_1 k_2 k_3 +k_1 k_2  k_4+ k_1 k_3 k_4 + k_2 k_3 k_4 )~.
\end{equation}
The effective cone for this manifold consists of an infinite number of additional K\"ahler cones, neighbouring the the four boundaries of $\cK(X)$, which corresponds to bi-rationally equivalent and isomorphic Calabi-Yau three-folds related to $X$ by sequences of flops (see Refs.~\cite{Brodie:2021toe}). These additional cones are obtained from the K\"ahler cone by the action of a group generated by
\begin{equation*}
M_1 {=} \left(\begin{array}{rrrr}{\!\!\!\!-1}&{0}&{0}&{0}\!\!\\{2}&{1}&{0}&{0}\!\!\\{2}&{0}&{1}&{0}\!\!\\{2}&{0}&{0}&{1}\!\!\end{array}\right)~,~
M_2 {=} \left(\begin{array}{rrrr}{\!\!1}&{2}&{0}&{0}\!\!\\{\!\!0}&{\!\!\!\!{-}1}&{0}&{0}\!\!\\{\!\!0}&{2}&{1}&{0}\!\!\\{\!\!0}&{2}&{0}&{1}\!\!\end{array}\right)~,~
M_3 {=} \left(\begin{array}{rrrr}{\!\!1}&{0}&{2}&{0}\!\!\\{\!\!0}&{1}&{2}&{0}\!\!\\{\!\!0}&{0}&{\!\!\!\!{-}1}&{0}\!\!\\{\!\!0}&{0}&{2}&{1}\!\!\end{array}\right)~,~
M_4 {=} \left(\begin{array}{rrrr}{\!\!1}&{0}&{0}&{2}\!\!\\{\!\!0}&{1}&{0}&{2}\!\!\\{\!\!0}&{0}&{1}&{2}\!\!\\{\!\!0}&{0}&{0}&{\!\!\!\!{-}1}\!\!\end{array}\right)~.
\end{equation*}

Consequently, any effective non-trivial line bundle $L$ is related to a line bundle $L'$ belonging to the closure of the K\"ahler cone by a finite number of transformations
\begin{equation}
c_1({L'})=M_{i_1}M_{i_2}\ldots M_{i_k}c_1(L) \in \overline{\cK(X)}~.
\end{equation}
Since the number of global sections of a line bundle is invariant under flops, it follows that 
\begin{equation}
h^0(X,L)=h^0(X,L')= \chi(X,L')\;, 
\end{equation}
where the Euler characteristic can be computed from Eq.~\eqref{eq:7862_index} and the second equality holds by Kodaira's vanishing theorem and the Kawamata-Viehweg vanishing theorem (needed on the walls separating the K\"ahler cone of $X$ from the neighbouring K\"ahler cones). In fact there are a number of two-faces of $\overline{\cK(X)}$ which do not belong to the interior of the extended K\"ahler cone and consequently are not covered by the Kawamata-Viehweg vanishing theorem. These correspond to two of the integers $k_i$ vanishing and the other two being non-negative, which we denote by $k_A$ and $k_B$. In these cases the zeroth cohomology function is simply $(1+k_A)(1+k_B)$, which can be easily traced back to the zeroth cohomology of two line bundles on $\mathbb P^1\times \mathbb P^1$. 

This procedure gives an extremely efficient method for computing the zeroth cohomology of line bundles on the tetra-quadric threefold. Alternatively, one could write down an explicit formula containing an infinite number of case distinctions corresponding to the infinite number of K\"ahler cones obtained by flopping~$X$. In practice, however, only a small number of such cohomology chambers matters, since the chambers are increasingly thing away from the original K\"ahler cone $\mathcal K(X)$ and contain line bundles where at least one of the integers $k_i$ is very large.

Once the zeroth cohomology is known, the third cohomology follows by Serre duality, 
\begin{equation}
h^3(X,L) = h^0(X,L^*)~.
\end{equation}
Note that since the effective cone is convex there are no line bundles, except for the trivial line bundle, that have both $h^0(X,L)$ and $h^3(X,L)$ non-vanishing.

The middle cohomologies are related to the zeroth and the third cohomologies by the formula
\begin{equation}\label{eq:EulerCh}
h^1(X,L)-h^2(X,L) = h^0(X,L)-h^3(X,L)-\chi(X,L)~.
\end{equation}
On the tetra-quadric manifold it turns out that almost all line bundles either have $h^1(X,L)=0$ or $h^2(X,L)=0$. In all these cases Eq.~\eqref{eq:EulerCh} provides a formula for the middle cohomologies. The exceptions correspond to line bundles for which two of the $k_i$ integers are zero and the other two have opposite sign and are greater than $1$ in modulus. If $k_A$ and $k_B$ denote the non-zero integers, it turns out that in all these exceptional cases the following simple relation holds
\begin{equation}
h^1(X,L)+h^2(X,L) = -2(1+k_A k_B)~,
\end{equation}
which together with Eq.~\eqref{eq:EulerCh} fixes the middle cohomologies. 

\subsection{The role of machine learning}
Machine learning played an important role in the initial identification of cohomology formulae~\cite{Klaewer:2018sfl, Brodie:2019dfx} on complex surfaces and threefolds. It has also been used in the context of line bundles over complex curves in Ref.~\cite{Bies:2020gvf}. 

More concretely, in Ref.~\cite{Brodie:2019dfx} it was shown that the standard black box approach based on simple fully connected networks is not of much use for the problem of finding analytic cohomology formulae, which requires the simultaneous learning of the chamber structure, as well as the polynomials describing the cohomology function in each chamber. Instead, a three-step procedure was shown to be successful. First, a neural network is set up for the purpose of learning the number of chambers and their approximate boundaries. For each so-obtained region, the corresponding cohomology polynomial can then be found by a simple fit. Finally, the polynomials are used to determine the exact boundaries of the cohomology chambers. The algorithm is capable of learning the piece-wise polynomial cohomology formulae. 

Conversely, and relying on the theoretical understanding of the structure of cohomology formulae~\cite{Brodie:2019ozt, Brodie:2020wkd, Brodie:2020fiq}, it was shown that machine learning of cohomology data can be used to derive information about the geometric properties of the manifold~\cite{Brodie:2019dfx}. For instance, in the case of complex projective surfaces, the information about the nef cone and the Mori cone is sufficient to determine all line bundle cohomologies. However, it can be hard to obtain this information, in general. On the other hand, algorithmic methods for computing line bundle cohomology can be employed to obtain enough training input as needed to learn the cohomology formulae and then use these to extract the information about the nef cone and the Mori cone. In the case of Calabi-Yau threefolds, machine learning of cohomology formulae can be used to extract information about flops, rigid divisors and Gromov-Witten invariants.

For string theory applications the cohomology formulae become useful if the entire chamber structure and the piecewise quasi-polynomial functions are known for both the zeroth and the higher cohomologies. This can be difficult since the number of cohomology chambers increases quickly with the Picard number of the manifold and in many cases is infinite. 
Finding the boundaries of the chambers and the cohomology functions can be non-trivial and machine learning may help where algebro-geometric methods become unmanageable. On the other hand, we are currently lacking a theoretical understanding of the higher line bundle cohomologies on Calabi-Yau threefolds and machine learning can provide important hints about the underlying structures. Finally, machine learning can be used to go beyond the case of abelian bundles to explore the existence of analytic formulae for the cohomology of non-abelian bundles. 

\section{ML techniques for the computation of physical couplings}\label{sec:Yuk}
The goal of deriving quantitative predictions from heterotic string models hinges on the resolution of three difficult problems: 
\begin{itemize}\itemsep0em
\item[{\itshape i)}] the computation of (moduli dependent) physical couplings, relying on the knowledge of the Calabi-Yau metric and the  hermitian Yang-Mills connection on the holomorphic vector bundle;
\item[{\itshape ii)}] moduli stabilisation, that is the problem of fixing the free parameters of the internal geometry; 
\item[{\itshape iii)}] supersymmetry breaking and the derivation of the resulting properties at the electroweak scale via renormalization group analysis. 
\end{itemize}
In the following discussion we will mainly focus on the first problem, where machine learning is expected to make the strongest impact. 
Numerical computations of Calabi-Yau metrics and the hermitian Yang-Mills connections for fixed values of the moduli have been performed in Refs.~\cite{Headrick:2005ch, Douglas:2006hz, Doran:2007zn, Headrick:2009jz, Douglas:2008es, Anderson:2010ke, Anderson:2011ed, Cui:2019uhy} and more recently in Refs.~\cite{Ashmore:2019wzb, Douglas:2020hpv, Anderson:2020hux, Jejjala:2020wcc, Ashmore:2021ohf, Larfors:2021pbb, Ashmore:2021rlc} using machine learning techniques. 
Most of these methods have been implemented on a case-by-case basis. However, in order to analyse a relatively large number of models, as expected to arise from the automated searches discussed above, a more systematic approach is needed. 

\subsection{Physical Yukawa couplings}
One of the key steps towards realistic particle physics from string theory is to find models with the correct Yukawa couplings. The calculation of four-dimensional physical Yukawa couplings from string theory is notoriously difficult and proceeds in three steps. First, the holomorphic Yukawa couplings, that is, the trilinear couplings in the superpotential of the form 
\begin{equation}\label{eq:Yukawa}
\lambda_{IJK}\propto \int_X \bar\Omega \wedge \nu_{I}^a\wedge \nu_{J}^b\wedge \nu_{K}^c  f_{abc}
\end{equation}
have to be determined.  Here $\Omega$ denotes the homomorphic $(3,0)$-form on the Calabi-Yau threefold~$X$, while $f_{abc}$ are structure constants descending from the structure constants of $E_8$. The 1-forms $\nu_I^a$, $\nu_J^b$ and $\nu_K^c$ correspond to the matter fields and are harmonic. However, the integral in Eq.~\eqref{eq:Yukawa} is quasi-topological and depends only on the cohomology classes of the 1-forms. This fact greatly simplifies the computation of the holomorphic Yukawa couplings, which can be accomplished either by algebraic methods~\cite{Candelas:1987se, Braun:2006me, Anderson:2009ge, Anderson:2010tc} or by methods rooted in differential geometry~\cite{Candelas:1987se,Blesneag:2015pvz, Blesneag:2016yag, Buchbinder:2016jqr}. Although non-trivial, these computations can in principle keep track analytically of the moduli dependence. 

The second step is the calculation of the matter field K\"ahler metric which determines the field normalisation and the re-scaling required to convert the holomorphic couplings into the physical Yukawa couplings. The matter field K\"ahler metric takes the form
\begin{equation}
 G_{IJ}\propto \int_X \nu_I\wedge \bar{\star}_V ({\nu}_J)\; , \label{G0}
\end{equation} 
where $\bar{\star}_V$ refers to a Hodge dual combined with a complex conjugation and an action of the hermitian bundle metric on $V$. This quantity is non-holomorphic and requires not only the harmonic representatives for the $1$-forms, but also  knowledge of the Ricci-flat metric on $X$ and the hermitian Yang-Mills connection on $V$ both of which enter in the definition of the Hodge dual~$\bar{\star}_V$. 
The third step in the computation of physical Yukawa couplings involves the stabilisation of the moduli. The existing methods are typically unable to fix all the moduli perturbatively, having to rely on difficult to handle non-perturbative arguments. A possible approach here could be to insert the values of the moduli stabilised at the perturbative level into the moduli-dependent numerical expressions for the physical Yukawa couplings and to use these values to infer in a bottom-up manner the required VEVs for the unstabilised moduli. This is a numerical optimisation problem where machine learning can once again make a difference.

\subsection{Calabi-Yau metrics and hermitian Yang-Mills connections}
The only class of heterotic Calabi-Yau models where an analytic expression for the matter field K\"ahler metric is known corresponds to standard embedding models. In this case, the matter field K\"ahler metrics for the $(1,1)$ and $(2,1)$ matter fields are essentially given by the metrics on the corresponding moduli spaces~\cite{Candelas:1987se, Candelas:1990pi}. 
For non-standard embeddings things are more complicated and, unfortunately, there are no known analytic expressions\footnote{Recently analytic expressions for K3 metrics have been found in Ref.~\cite{Kachru:2020tat}, however the methods used there do not have an immediate generalisation to threefolds.}  for  the Ricci-flat metric on $X$ and the hermitian Yang-Mills connection on $V$, except for certain approximations in a number of special cases \cite{Blesneag:2018ygh}.  One approach is to use Donaldson's numerical algorithm to determine the Ricci-flat Calabi-Yau metric~\cite{Donaldson:2001, Donaldson:2005, DonaldsonNumerical} and the subsequent work applying this algorithm to various explicit examples and to the numerical calculation of the Hermitian Yang-Mills connection on vector bundles~\cite{Wang:2005, Headrick:2005ch, Douglas:2006hz, Doran:2007zn, Headrick:2009jz, Douglas:2008es, Anderson:2010ke, Anderson:2011ed}. 
A significant drawback of this method is that it provides numerical expressions for the required quantities only at fixed values of the moduli; trying different points in the moduli space corresponds to re-running the algorithm from scratch which can be computationally very intensive. 
\vspace{8pt}

{\bfseries Ricci-flat metrics on Calabi-Yau threefolds.}
Following the work of Calabi and Yau we known that every compact K\"ahler manifold $X$ with vanishing first Chern class has a unique Ricci-flat metric in every K\"ahler class. The problem of finding a metric $g_{\rm CY}$ with vanishing Ricci curvature can be simplified to the problem of finding a metric with a prescribed volume form. Thus, if $J_{\rm CY}$ denotes the K\"ahler form associated with the unique Ricci-flat metric in a given class $[J_{\rm CY}]$, it can be shown that $J_{\rm CY}$ must satisfy the equation
\begin{equation}\label{eq:CY}
J_{\rm CY}\wedge J_{\rm CY}\wedge J_{\rm CY} = \kappa\, \Omega\wedge\bar\Omega~,
\end{equation}
for a certain number $\kappa\in\mathbb C$ that only depends on the moduli. In order to find $J_{\rm CY}$ one can start with a K\"ahler form $J'$ in the same cohomology class, which must be related to $J_{\rm CY}$ by
\begin{equation}
J_{\rm CY}=J' + \partial\bar\partial \phi~,
\end{equation}
for some smooth zero-form $\phi$ on $X$. Thus the problem of finding the Ricci-flat metric boils down to finding the zero-form $\phi$ that satisfies Eq.~\eqref{eq:CY} -- this is the Monge-Ampere equation for which Yau's non-constructive proof showed that a solution must exist \cite{Yau:1978}. The simplification brought by this reformulation is important, since the Ricci curvature depends on the second derivatives of the metric, while Eq.~\eqref{eq:CY} involves only the metric and not its derivatives. 

There have been several proposals for how to train a neural network in order to learn the Calabi-Yau metric. Here we outline a direct method of learning the metric, as used in Ref.~\cite{Larfors:2021pbb} for the case of Calabi-Yau threefolds constructed as complete intersections in products of projective spaces. In this case one can start with the K\"ahler form $J'$ given by the pull-back to~$X$ of the Fubini-Study form on the embedding space. The process of finding $\phi$ then proceeds in a self-supervised learning fashion by uniformly sampling points of $X$ and minimising a loss function that takes into account: 
\begin{itemize}\itemsep0em
\item[$i)$] how well the Monge-Ampere equation is satisfied; 
\item[$ii)$] the amount by which the form $J'+\partial\bar\partial\phi$ fails to be closed; 
\item[$iii)$] the amount by which different expressions fail to match on overlapping patches;
\item[$iv)$] the amount by which the class of $J'+\partial\bar\partial\phi$ deviates from the original class of $J'$, as measured by the corresponding overall volumes;
\item[$v)$] the amount by which the Ricci curvature fails to vanish.
\end{itemize}
The computation of the Ricci-loss is expensive, as it involves derivatives of the metric. In fact, this loss is not needed, as it is already taken into account by the Monge-Ampere loss, however, it can be used as a cross-check. A key advantage of using neural networks is that numerical metrics can be computed relatively quickly (a few hours on a laptop) for any values of the moduli. 
\vspace{12pt}

{\bfseries Hermitian Yang-Mills connections on holomorphic line bundles.} 
Solving the hermitian Yang-Mills euqation $g^{a\bar b}F_{a\bar b}=0$ requires the use of a previously trained network to compute the Ricci-flat metric $g^{a\bar b}$. Provided that such a neural network exists, the training of the connection network can proceed in a similar self-supervised fashion, by sampling a large number of points on the manifold and minimising a loss function that takes into account the amount by which $g^{a\bar b}F_{a\bar b}$ fails to vanish as well as the gluing conditions between patches. Initial steps in this direction have been taken in Ref.~\cite{Ashmore:2021rlc}. 

The study of harmonic forms, needed for the computation of the matter field K\"ahler metric, also boils down to finding numerical solutions to PDEs, in this case Laplace's equation on Calabi-Yau threefolds, and can be approached using similar self-supervised methods (see Refs.~\cite{Ashmore:2020ujw, Ashmore:2021qdf} for some recent work).

\section{Conclusions}
The primary message of this review is that the ongoing developments in string phenomenology and the new opportunities opened up by machine learning make the problem of embedding the Standard Model of particle physics into string theory much more likely to be resolved in the near future. If successful, this monumental effort would provide an ultraviolet completion of particle physics and a natural setting to address the physics beyond the Standard Model, including quantum gravity.
The incorporation of AI tools into string theory make possible the implementation of an unprecedented scrutiny of the string landscape and facilitate the derivation of numerical values for the physical couplings in realistic string models. The resolution of these long standing issues in string phenomenology would represent a major advancement in fundamental physics, with the prospect of deriving from first principles fundamental quantities in nature, such as the mass of the electron.

\section*{Acknowledgements}
My work is supported by a Stephen Hawking Fellowship, EPSRC grant EP/T016280/1.

\end{document}